# Людина з комп'ютерним обличчям
## (до 80-річчя Айвена Едварда Сазерленда)

Сергій Олексійович Семеріков
Кафедра інформатики та прикладної математики,
Криворізький державний педагогічний університет,
пр. Гагаріна, 54, м. Кривий Ріг, 50086, Україна
semerikov@gmail.com

Андрій Миколайович Стрюк[*], Катерина Іванівна Словак[ǂ],
Наталя Василівна Рашевська[#], Юлія Володимирівна Єчкало[¶]
Кафедра моделювання та програмного забезпечення[*], кафедра вищої математики[##], кафедра фізики[¶], ДВНЗ «Криворізький національний університет», вул. Віталія Матусевича, 11, м. Кривий Ріг, 50027, Україна
andrey.n.stryuk@gmail.com[*], slovak@fsgd.ccjournals.eu[ǂ],
nvr1701@gmail.com[#], uliaechk@gmail.com[¶]

**Анотація**. У статті наведено основні віхи науково-технологічної біографії Айвена Едварда Сазерленда. Показано вплив сім'ї та школи на розвиток його дослідницьких компетентностей, наведено маловідомі біографічні факти, що пояснюють еволюцію його наукових інтересів: від динамічних об'єктно-орієнтованих графічних систем через системи віртуальної реальності до асинхронної логіки.

**Ключові слова**: Айвен Едвард Сазерленд; Sketchpad; віртуальна реальність; асинхронна логіка.

**S. O. Semerikov**[±], **A. M. Struik**[*], **K. I. Slovak**[ǂ], **N. V. Rashevska**[#], **Yu. V. Yechkalo**[¶]. **A man with a computer face (to the 80th anniversary of Ivan Edward Sutherland)**

**Abstract**. The article presents the main milestones of the science and technology biography of Ivan Edward Sutherland. The influence of the family and the school on the development of its research competencies is shown, and little-known biographical facts explaining the evolution of his scientific interests is presented: from dynamic object-oriented graphic systems through systems of virtual reality to asynchronous circuits.

**Keywords**: Ivan Edward Sutherland; Sketchpad; virtual reality; asynchronous circuits.

**Affiliation**: Department of Computer Science and Applied Mathematics, Kryvyi Rih State Pedagogical University, 54, Gagarin Ave., Kryvyi Rih, 50086, Ukraine[±];

Department of simulation and software[*], Department of higher mathematics[##], Department of physics[¶], SIHE «Kryvyi Rih National

University», 11, Vitalyy Matusevych Str., Kryvyi Rih, 50027, Ukraine.
E-mail: semerikov@gmail.com±, andrey.n.stryuk@gmail.com*, slovak@fsgd.ccjournals.eu‡, nvr1701@gmail.com#, uliaechk@gmail.com¶.

Айвен Едвард Сазерленд (Ivan Edward Sutherland) народився 16 травня 1938 р. у м. Гастінгс (штат Небраска, США). Його батько був доктором філософії (PhD) із цивільної інженерії, а мати – шкільним учителем, і це сприяло розвитку інтересу до навчання у нього та його старшого брата Берта (William Robert "Bert" Sutherland).

Його першим комп'ютером був Simon – релейний електромеханічний комп'ютер, створеним Е. К. Берклі (Edmund Callis Berkeley). Призначений скоріше для навчальної мети – демонстрації концепції цифрових обчислень, – Simon не міг бути використаний для розв'язання скільки-небудь значних задач через апаратні обмеження:

– реєстри та АЛП зберігали лише 2 біти;

– користувач міг використовувати для введення даних лише п'ять клавіш на передній панелі;

– результат відображався за допомогою п'яти ламп.

Програми Simon виконувалися зі стандартної паперової стрічки з п'ятьма рядками отворів для даних. Перфострічка слугувала не тільки для введення даних, але і як пам'ять для зберігання (рис. 1). Комп'ютер міг виконувати чотири операції: додавання, заперечення, більше та вибір. Тому перша велика комп'ютерна програма Айвена реалізувала на Simon операцію ділення. Щоб зробити ділення можливим, він за допомогою брата додав до набору інструкцій Simon умовну зупинку. Комп'ютер був запрограмований за допомогою перфострічки, а алгоритм ділення був найдовшою програмою, яку коли-небудь писали для Simon – це була паперова стрічка довжиною близько 2,5 метрів.

«Одного дня ми зможемо навіть мати маленькі комп'ютери у наших будинках, що будуть живитися від ліній електропередач, як холодильники або радіо ... Вони зможуть нагадати нам факти, які нам буде важко пригадати. Вони зможуть розрахувати рахунки та податки на прибуток. Школярі з домашнім завданням зможуть звернутися до них за допомогою. Вони зможуть навіть пробігти та перелічити комбінації можливостей, які нам потрібно враховувати при прийнятті важливих рішень» [1, с. 42].

У 1955 році А. Е. Сазерленд закінчив Скарсдейльську старшу школу (м. Скарсдейл, штат Нью-Йорк).

Відповідно до власного резюме від 25 липня 2017 року [4], у 1959 році в Технологічному інституті Карнегі (сьогодні – Університет Карнегі-Меллона, м. Піттсбург, штат Пенсильванія) здобув ступінь бакалавра, у

1960 – ступінь магістра у Каліфорнійському технологічному інституті (м. Пасадена, штат Каліфорнія), а у 1963 році – ступінь доктора філософії у Массачусетському технологічному інституті (м. Кембрідж, штат Массачусетс); всі – у галузі електричної інженерії [7]. Під час навчання з 1955 по 1959 рік отримував стипендію імені Джорджа Вестінгауза, двічі – у 1958 та 1959 рр. – був переможцем конкурсу студентських наукових робіт (American Institute of Electrical Engineers Student Prize Paper Contest for District 2 Winner), а з 1959 по 1962 рік отримував стипендію Національного наукового фонду (National Science Foundation Fellowship).

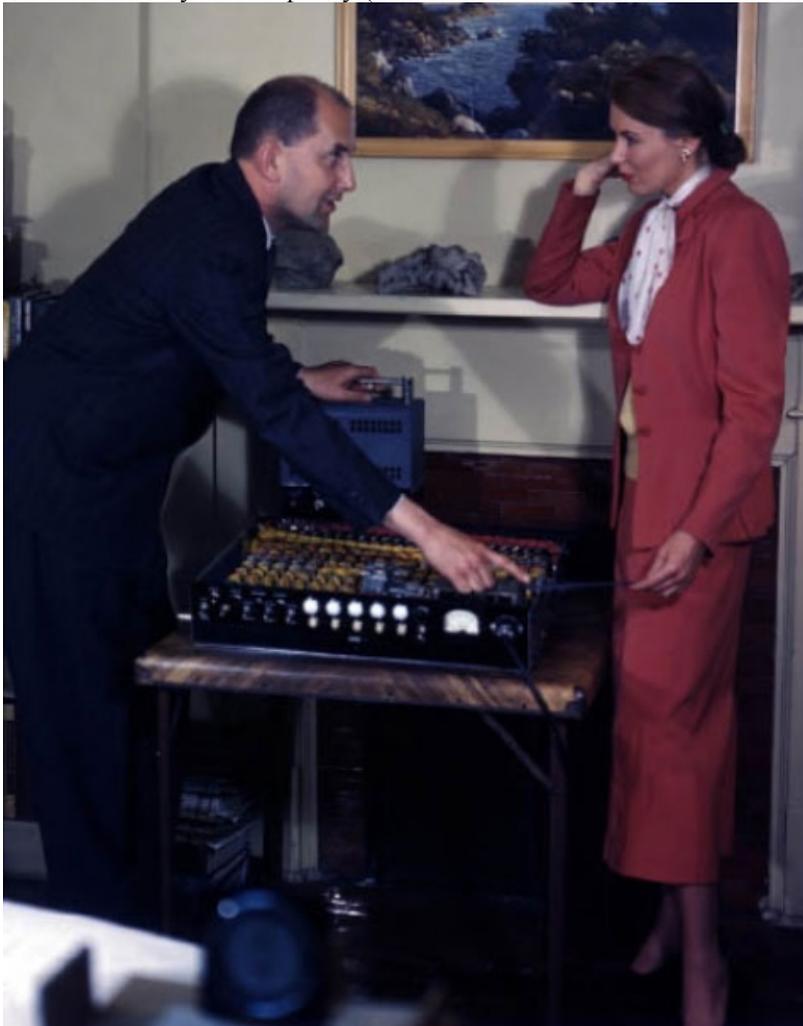

Рис. 1. Едмунд Берклі пояснює, як Simon зчитує команди з перфострічки

А. Е. Сазерленд є членом Асоціації обчислювальної техніки (Association for Computing Machinery Fellow), Національної академії інженерії (National Academy of Engineering member), Національної академії наук США (Member, United States National Academy of Sciences), Американської асоціації сприяння розвитку науки (American Association for the Advancement of Science), Інституту інженерів з електротехніки та електроніки (IEEE – Institute of Electrical and Electronics Engineers).

А. Е. Сазерленд має почесні ступені:

– магістра мистецтв (1966 рік, Гарвардський університет, м. Кембрідж, штат Массачусетс);

– доктора наук (1986 рік, Університет Північної Кароліни, м. Чепел Хілл, штат Північна Кароліна);

– доктора філософії (2000 рік, Університет Юти, м. Солт Лейк Сіті, штат Юта; 2003 рік, Університет Карнегі-Меллона, м. Піттсбург, штат Пенсільванія).

Також відзначений рядом нагород:

1972 рік – нагорода імені Зворикіна Національної академії інженерії (National Academy of Engineering First Zworykin Award);

1983 рік – премія Стівена Е. Кунса Асоціації обчислювальної техніки (SIGGRAPH);

1986 рік – премія Емануеля Р. Піора від IEEE (IEEE Emanuel R. Piore Award) за піонерський внесок у розвиток інтерактивних комп'ютерних графічних систем та внесок в інформатичну освіту;

1987 рік – нагорода за лідерство почесної програми Computerworld (Computerworld Honors Program, Leadership Award);

1988 рік – премія Тьюрінга Асоціації обчислювальної техніки (ACM Turing Award) за інноваційний і далекоглядний внесок до комп'ютерної графіки, починаючи зі Sketchpad та продовжуючи далі;

1993 рік – нагорода у галузі програмних систем Асоціації обчислювальної техніки (ACM Software System Award) за Sketchpad;

1994 рік – нагорода першопрохідця від Фонду електронних рубежів (Electronic Frontier Foundation EFF Pioneer Award);

1995 рік – нагорода піонера віртуальної реальності від CyberEdge (рис. 2);

1996 рік – сертифікат «За заслуги» Інституту Франкліна (Franklin Institute's Certificate of Merit), нагорода Price Waterhouse за визначні досягнення в галузі інформаційних технологій (Price Waterhouse Information Technology Leadership Award for Lifetime Achievement);

1998 рік – медаль Джона фон Неймана від IEEE (IEEE John von Neumann Medal);

2005 рік – почесний член Музею комп'ютерної історії (Computer

History Museum Fellow) за CAD Sketchpad і внесок у комп'ютерну графіку та освіту;

2012 рік – премія Кіото (Kyoto Prize) у категорії передових технологій (інформатика) за піонерські досягнення у розвитку комп'ютерної графіки та інтерактивних інтерфейсів (рис. 3);

2016 рік – уведений до Національного Залу слави винахідників за відображення вікон шляхом відсікання [3].

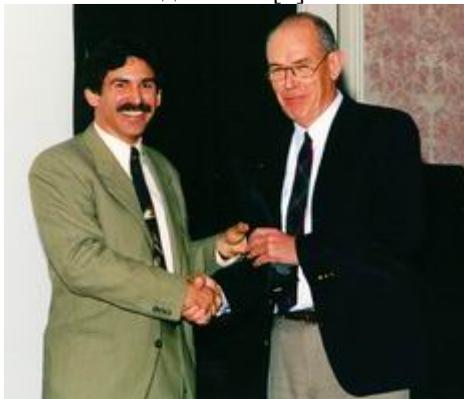

Рис. 2. А. Е. Сазерленд приймає нагороду піонера віртуальної реальності від CyberEdge

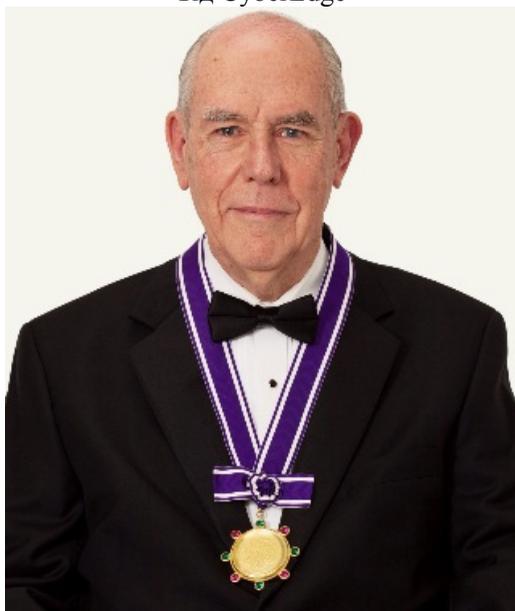

Рис. 3. А. Е. Сазерленд із медаллю премії Кіото

Із неповного переліку нагород А. Е. Сазерленда щонайменше у двох відзначено Sketchpad («Етюдник», також відомий як Robot Draftsman – «Робот-кресляр») – найбільш відома рання реалізація віртуальної реальності та одночасно – батько CAD-систем, створений у рамках його докторської дисертації під керівництвом К. Е. Шеннона (Claude Elwood Shannon).

На той час для проекту військово-повітряних сил SAGE були потрібні швидкі комп'ютери з можливостями графічного введення та виведення. Світлове перо було спеціально винайдено для проекту SAGE, але комп'ютерні дисплеї залишались грубою адаптацією радарних. Основний комп'ютер проекту TX2, завершений у 1959 році, був не лише повністю транзисторним, але й мав масив пристроїв для введення та виведення в режимі реального часу: світлове перо, 16 слів (39-бітних) перемикачів, які можна було протестувати, 4 цифрові ручки управління, мікрофон та динаміки. Його графічний дисплей був дев'ятидюймовим пристроєм прямого доступу із роздільною здатністю 1024 × 1024 пікселів на 9 дюймів. Не дивно, що студенти-випускники використали TX2 як платформу для створення перших відеоігор – Tic-Tac-Toe, Mouse in the Maze та Spacewar.

Дисертація А. Е. Сазерленда на здобуття наукового ступеня доктора філософії отримала назву «Етюдник, людино-машинна графічна комунікаційна система» [10]. Sketchpad може виглядати очевидною програмою для малювання з сьогоднішньої перспективи, але нічого подібного до нього просто не було. Sketchpad вплинув на всі форми взаємодії людини з комп'ютером. Так, він передбачав активне використання світлового пера для малювання та маніпулювання графічними об'єктами. Sketchpad міг приймати обмеження та задавати взаємозв'язки між сегментами та дугами, малювати горизонтальні та вертикальні лінії, об'єднувати їх у фігури, які можна було копіювати, переміщувати, повертати та масштабувати, зберігаючи їх основні властивості. У Sketchpad вперше було використано віконний інтерфейс та алгоритм відсікання вікон, що дозволяв масштабування (рис. 4).

Коли А. Е. Сазерленда спитали, як він зміг створити першу інтерактивну графічну програму, першу непроцедурну мову програмування, першу об'єктно-орієнтовану програмну систему за один рік, він відповів: «Ну, я ж не знав, що це важко» [5].

Під час роботи над дисертацією у 1960-1962 рр. А. Е. Сазерленд працював у дослідницькій лабораторії електроніки Массачусетського технологічного інституту та лабораторії Лінкольна, спільної із Міністерством оборони США (в останній – улітку). Саме тому, коли з 25 лютого 1963 року по 25 лютого 1965 року А. Е. Сазерленд перебував на військовій службі в армії США, він проходив службу на посаді першого лейтенанта Сигнального корпусу спочатку в Мічиганському університеті, а далі в Агентстві національної безпеки США. Перебуваючи на службі, у 1964 році він змінив Дж. К. Р. Ліклайдера (Joseph Carl Robnett Licklider) на посаді голови офісу технологій опрацювання інформації (Information Processing Technology Office) Агенції передових дослідницьких проектів (Advanced Research Projects Agency, зараз відомої як DARPA) Міністерства оборони США (працював до 1966 року).

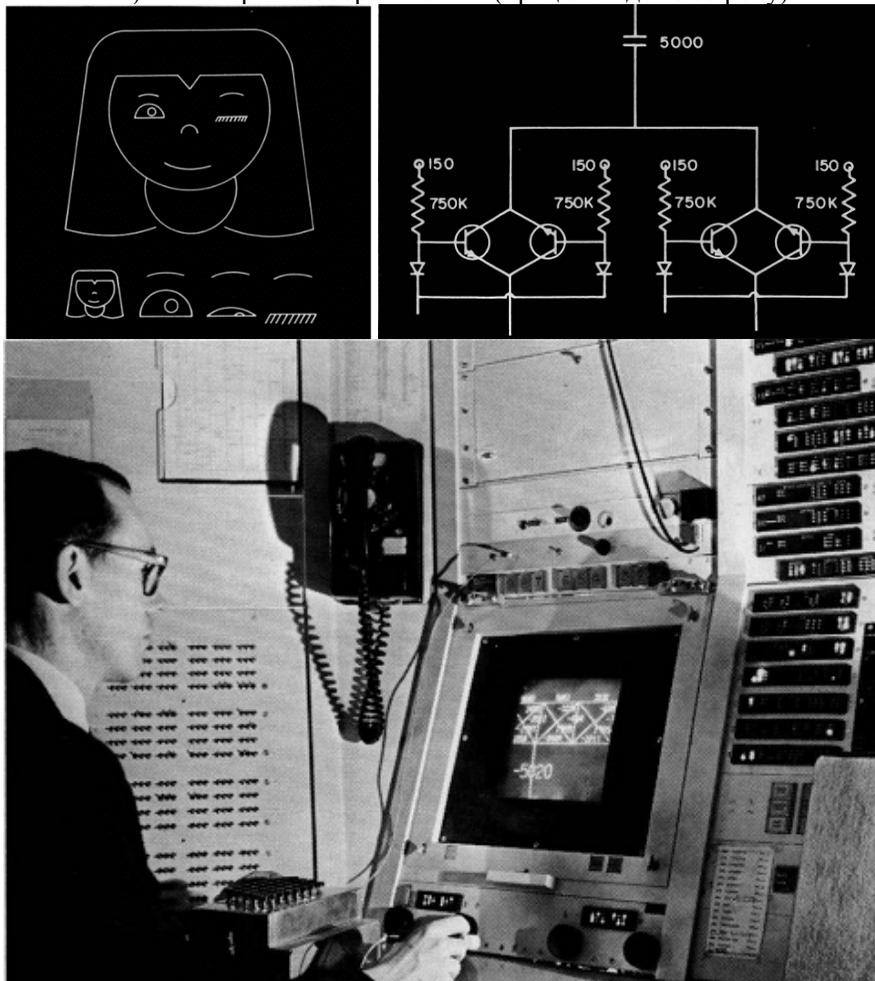

Рис. 4. А. Е. Сазерленд демонструє роботу із графічними об'єктами у Sketchpad

У 1965 році А. Е. Сазерленд увів концепцію ідеального (остаточного) дисплею (ultimate display), з'єднаного із комп'ютером для полегшення знайомства із поняттями, які неможливо реалізувати у фізичному світі: «Ідеальним дисплеєм буде, звичайно, кімната, всередині якої комп'ютер зможе контролювати існування матерії. Крісло, зображене в такій кімнаті, буде достатньо зручними, щоб на ньому сидіти. Наручники, зображені в такій кімнаті, будуть стримувати, а куля, зображена в такій кімнаті, буде фатальною. За відповідного програмування такий дисплей міг буквально стати Країною Чудес, до якої подорожувала Аліса» [12, с. 507-508]. Опис дисплею, наведений А. Е. Сазерлендом, включає як візуальні, так і кінестетичні стимули. Останнє стимулювало Ф. П. Брукса (Frederick Phillips "Fred" Brooks Jr.) розпочати у 1967 році в Університеті Північної Кароліни проект GROPE для дослідження використання кінестетичної взаємодії як засобу, що допомагає біохімікам «відчути» взаємодію між протеїновими молекулами (рис. 5). Свій варіант дисплею Ф. П. Брукс назвав гаптичним – таким, що надає відчуття (дотик, температура, тиск тощо), опосередковані шкірою, м'язами, сухожиллями або суглобами [2, с. 177-178].

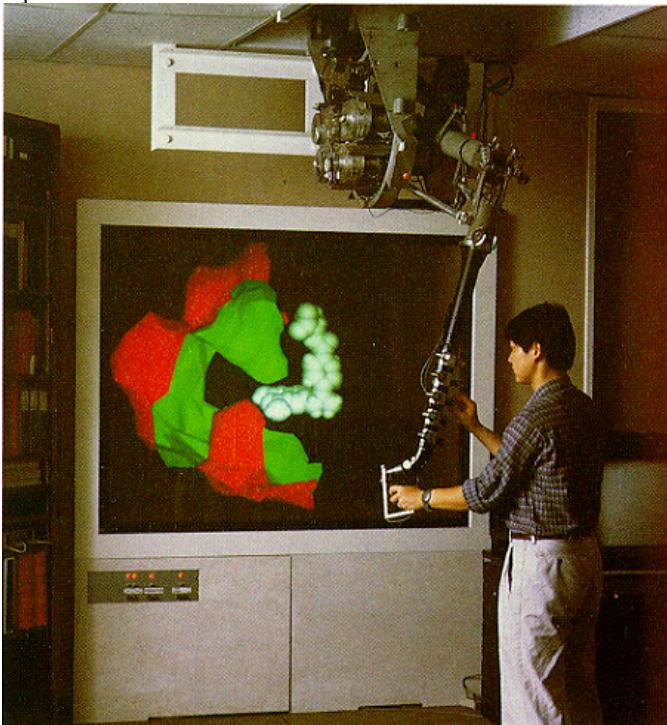

Рис. 5. Гаптичний дисплей GROPE-III

З 1965 року по 1968 рік А. Е. Сазерленд працював у Гарвардському університеті на посаді асоційованого професора електричної інженерії, де за допомогою свого студента Р. Ф. Спрула (Robert Fletcher "Bob" Sproull) у 1968 році він створив перший варіант ідеального дисплею – шолом (head-mounted display) доповненої реальності: поєднання фізичних та цифрових просторів у семантично пов'язаних контекстах, для яких об'єкти асоціацій розташовані у реальному світі. На відміну від віртуальної реальності, доповнена не створює повністю віртуальне середовище, а поєднує віртуальні елементи з реальним світом: до реального оточення користувача додаються віртуальні об'єкти, що змінюються унаслідок його дій. А. Е. Сазерленд у роботі [9] вказує, що це вимагає створення віртуальних інструментів або компонентів, керованих користувачем, для виконання певних досліджень, проведення експерименту тощо. Розроблений ним шолом віртуальної та доповненої реальності має влучну назву «Дамоклів меч» (The Sword of Damocles) – через велику вагу та розміри механізм були стаціонарно змонтований над користувачем (рис. 6). Середовище доповненої реальності створювалось шляхом накладання простих комп'ютерних моделей на зображення реального світу (рис. 7).

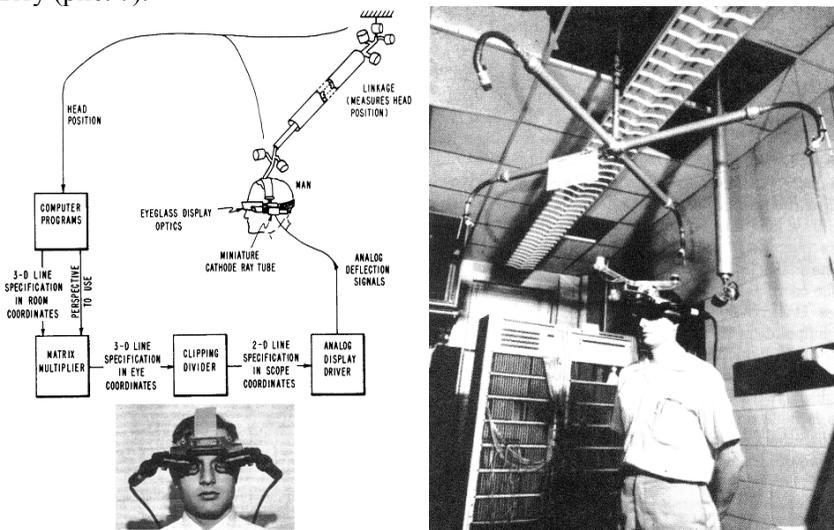

Рис. 6. Загальна схема роботи та зовнішній вигляд «Дамоклового меча» А. Е. Сазерленда та Р. Ф. Спрула [9, с. 296-298]

Інший з його студентів, Д. Коен (Danny Cohen), був першим, хто розробив візуальні симулятори польоту та радара. Робота на авіасимуляторі привела до розробки алгоритму Коена-Сазерленда для тривимірного відсікання ліній [8].

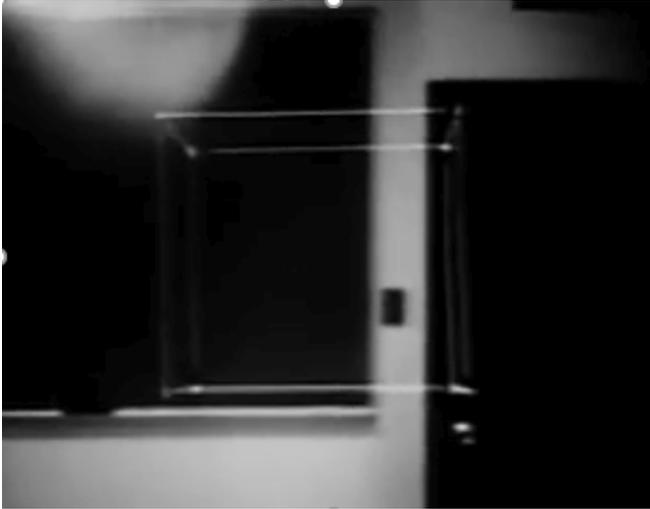

Рис. 7. Комп'ютерна модель куба, що доповнює оточуюче середовище
(кадр, знятий через камеру «Дамоклового меча»)

У 1968 році А. Е. Сазерленд став співзасновником та віце-президентом Evans and Sutherland Computer Corporation, об'єднавшись із Д. Евансом (David Evans) для створення центру досліджень комп'ютерної графіки при Університеті Юти, де з 1968 по 1974 рік А. Е. Сазерленд працював професором комп'ютерних наук. Компанія проводила піонерську роботу в галузі апаратного забезпечення для прискорення тривимірної графіки реального часу та створення принтерних мов. Серед колишніх працівників Evans and Sutherland Computer Corporation – майбутні засновники Adobe Дж. Е. Варнок (John Edward Warnock) та Silicon Graphics Дж. Г. Кларк (James Henry Clark).

«Дослідження – це весело! Як і в командному спорті, полювання на нові знання дає мету, товариство, розмову, змагання та визнання. Пошук нових знань приносить радість новизни, краси, простоти, розуміння, а іноді – й корисності. ... Ми всі повинні пам'ятати, що дослідження – це людські зусилля, яким загрожують технічні та емоційні ризики і розчарування. Скоротіть важку роботу, викоріньте розчарування, надихніть, надайте підтримку та визнайте досягнення, щоб досягти лояльності та результатів» [15]. Серед його студентів того часу були: винахідник першої об'єктно-орієнтованої мови програмування Smalltalk А. К. Кей (Alan Curtis Kay), автор методу тонування поверхонь А. Гуро (Henri Gouraud), розробник перших практичних методів екранного згладжування Ф. Кроу (Franklin C. Crow), відкривач текстур та

В-сплайнів Е. Е. Кетмелл (Edwin Earl Catmull), розобник першої реалістичної 3D-анімації людського обличчя Ф. А. Парк (Frederic Ira Parke).

З 1974 по 1978 рік А. Е. Сазерленд працював Флетчерівським професором комп'ютерних наук та завідувачем відповідної кафедри Каліфорнійського технологічного інституту.

З 1974 року – старший технічний співробітник у RAND Corporation.

У 1980 році А. Е. Сазерленд став співзасновником, віце-президентом та технічним директором Sutherland, Sproull and Associates, Inc. У 1990 році вона була придбана компанією Sun Microsystems, Inc, де стала ядром їх дослідницького відділу – Sun Labs, а сам А. Е. Сазерленд – Sun Fellow, співробітником Sun найвищого технічного рангу (з 1991 року – віце-президентом Sun).

У 2005-2008 рр. А. Е. Сазерленд був запрошеним викладачем відділення комп'ютерних наук Каліфорнійського університету (м. Берклі, штат Каліфорнія).

Сьогодні А. Е. Сазерленд керує невеликою групою, що працює над асинхронними системами; його група розробляє методи асинхронної логіки та технології проектування для швидких схем КМОП і застосовує їх до нових апаратних архітектур. Його книга «Логічні зусилля», опублікована в 1999 році у співавторстві із Р. Ф. Спрулом та Д. Харрісом, описує математичні основи проектування швидких схем [13].

А. Е. Сазерленд – автор більше 60 патентів США, одним із найбільш цитованих з яких є [14]: система для поділу тривимірних багатокутників на класи з метою визначення видимих частин багатокутників для визначення тіней і ділянок багатокутників, обгороджених напівпрозорими поверхнями, і для визначення поліедрального взаємопроникнення (коли форми взаємоперетинаються). Серед інших патентів А. Е. Сазерленда найбільш значущими є:

– 7,994,501 Method and apparatus for electronically aligning capacitively coupled mini-bars;

– 7,786,427 Proximity optical memory module having an electrical-to-optical and optical-to-electrical converter;

– 7,660,842 Method and apparatus for performing a carry-save division operation;

– 7,636,361 Apparatus and method for high-throughput asynchronous communication with flow control;

– 7,417,993 Apparatus and method for high-throughput asynchronous communication;

– 7,384,804 Method and apparatus for electronically aligning capacitively coupled mini-bars;

– 7,200,830 Enhanced electrically-aligned proximity communication;
– 7,064,583 Arbiters with preferential enables for asynchronous circuits;
– 7,046,017 Full-wave rectifier for capacitance measurements;
– 7,026,867 Floating input amplifier for capacitively coupled communication;
– 7,020,779 Secure, distributed e-mail system;
– 7,012,459 Method and apparatus for regulating heat in an asynchronous system;
– 6,995,039 Method and apparatus for electrostatically aligning integrated circuits;
– 6,987,412 Sense amplifying latch with low swing feedback;
– 6,987,394 Full-wave rectifier for capacitance measurements;
– 6,925,411 Method and apparatus for aligning semiconductor chips using an actively driven vernier;
– 6,870,271 Integrated circuit assembly module that supports capacitive communication between semiconductor dies;
– 6,847,247 Jittery polyphase clock;
– 6,825,708 Apparatus and method for an offset-correcting sense amplifier;
– 6,812,046 Method and apparatus for electronically aligning capacitively coupled chip pads;
– 6,769,007 Adder circuit with a regular structure;
– 6,753,726 Apparatus and method for an offset-correcting sense amplifier;
– 6,741,616 Switch fabric for asynchronously transferring data within a circuit;
– 6,710,436 Method and apparatus for electrostatically aligning integrated circuits;
– 6,707,317 Method and apparatus for asynchronously controlling domino logic gates;
– 6,675,246 Sharing arbiter;
– 6,600,325 Method and apparatus for probing an integrated circuit through capacitive coupling;
– 6,559,531 Face to face chips;
– 6,500,696 Face to face chip;
– 6,496,359 Tile array computers;
– 6,486,709 Distributing data to multiple destinations within an asynchronous circuit;
– 6,456,136 Method and apparatus for latching data within a digital system;
– 6,420,907 Method and apparatus for asynchronously controlling state information within a circuit;
– 6,360,288 Method and modules for control of pipelines carrying data using pipelines carrying control signals;

– 6,356,117 Asynchronously controlling data transfers within a circuit;
– 6,351,261 System and method for a virtual reality system having a frame buffer that stores a plurality of view points that can be selected and viewed by the user;
– 6,304,125 Method for generating and distribution of polyphase clock signals;
– 6,188,262 Synchronous polyphase clock distribution system;
– 6,085,316 Layered counterflow pipeline processor with anticipatory control;
– 6,072,805 Observing arbiter;
– 5,955,898 Selector and decision wait using pass gate XOR;
– 5,943,491 Control circuit of mutual exclusion elements;
– 5,861,762 Inverse toggle XOR and XNOR circuit;
– 5,838,939 Multi-issue/plural counterflow pipeline processor;
– 5,805,838 Fast arbiter with decision storage;
– 5,758,139 Control chains for controlling data flow in interlocked data path circuits;
– 5,748,539 Recursive multi-channel interface;
– 5,742,182 Symmetric selector circuit for event logic;
– 5,713,025 Asynchronous arbiter using multiple arbiter elements to enhance speed;
– 5,684,724 Flashback simulator;
– 5,638,009 Three conductor asynchronous signaling;
– 5,600,848 Counterflow pipeline processor with instructions flowing in a first direction and instruction results flowing in the reverse direction;
– 5,592,103 System for fast switching of time critical input signals;
– 5,572,690 Cascaded multistage counterflow pipeline processor for carrying distinct data in two opposite directions;
– 5,567,110 Robot arm structure;
– 5,187,800 Asynchronous pipelined data processing system;
– 4,900,218 Robot arm structure;
– 4,679,213 Asynchronous queue system;
– 4,622,992 Reaction control valve;
– 4,209,240 Reticle exposure apparatus and method.

З 2009 року А. Е. Сазерленд працює в Коледжі інженерії та комп'ютерних наук імені Масіха Портлендського державного університету на посаді запрошеного професора та керує Центром асинхронних досліджень (Asynchronous Research Center) разом зі своєю дружиною Марлі Ронкен (Maria Elisabeth (Marly) Roncken). Сьогодні вони працюють над удосконаленням технології проектування автоматизованих та асинхронних схем, систем, методів та інструментів проектування (рис. 8) [6].

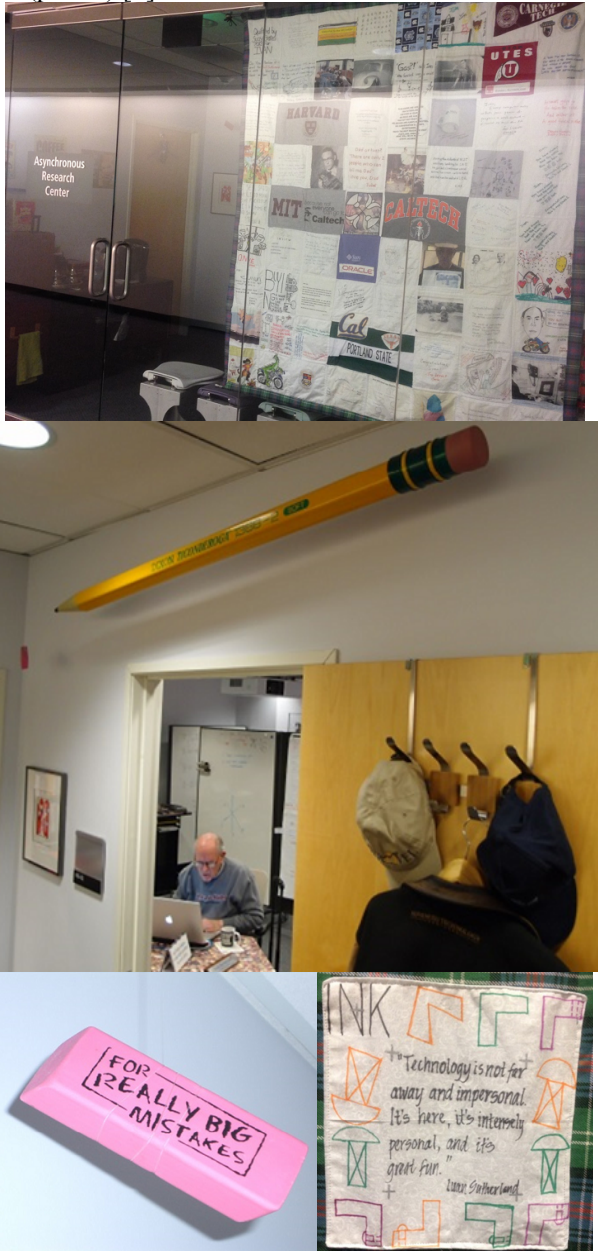

Рис. 8. А. Е. Сазерленд та його девіз у Центрі асинхронних досліджень Портлендського державного університету